\documentclass[aps,pra,twocolumn,groupedaddress,amsmath,amssymb,superscriptaddress,reprint,showpacs,showkeys,nofootinbib]{revtex4-2}
\usepackage{graphicx}
\usepackage{dcolumn}
\usepackage{color}
\usepackage{epsfig}
\usepackage{adjustbox}
\usepackage{subfigure}
\usepackage{amsmath}
\usepackage{multirow}
\usepackage{bm}
\usepackage{amsthm}
\usepackage{amsfonts}
\usepackage{float}
\usepackage{times}
\usepackage{amssymb}
\usepackage{relsize}
\usepackage{booktabs}
\usepackage[normalem]{ulem}
 
\usepackage[colorlinks=true,citecolor=blue,linkcolor=blue,urlcolor=blue]{hyperref}

\usepackage[sort&compress]{natbib}

\begin{document}
	
\title{An approach to interfacing the brain with quantum computers: practical steps and caveats}

\author{E. R. Miranda}
\email[Email:]{\qquad eduardo.miranda@plymouth.ac.uk}
\affiliation{ICCMR, University of Plymouth, Plymouth, UK}

\author{S. Venkatesh}
\affiliation{ICCMR, University of Plymouth, Plymouth, UK}

\author{J. D. Martin-Guerrero}
\author{C. Hernani-Morales}
\affiliation{IDAL, Electronic Engineering Department, ETSE-UV, University of Valencia, Valencia, Spain}

\author{L. Lamata}
\affiliation{Department of Atomic, Molecular and Nuclear Physics, University of Seville, Seville, Spain}

\author{E. Solano}
\affiliation{QuArtist, Physics Department, Shanghai University, Shanghai, China}
\affiliation{Ikerbasque, Bilbao, Spain}
\affiliation{Kipu Quantum, Munich, Germany}

\date{\today}

\begin{abstract}
We report on the first proof-of-concept system demonstrating how one can control a qubit with mental activity.  We developed a method to encode neural correlates of mental activity as instructions for a quantum computer. Brain signals are detected utilising electrodes placed on the scalp of a person, who learns how to produce the required mental activity to issue instructions to rotate and measure a qubit. Currently, our proof-of-concept runs on a software simulation of a quantum computer. At the time of writing, available quantum computing hardware and brain activity sensing technology are not sufficiently developed for real-time control of quantum states with the brain. But we are one step closer to interfacing the brain with real quantum machines, as improvements in hardware technology at both fronts become available in time to come. The paper ends with a discussion on some of the challenging problems that need to be addressed before we can interface the brain with quantum hardware.
\end{abstract}

\maketitle

\section{Introduction}

In a recent perspective paper ~\cite{QBraiNs2021}, we proposed the concept of Quantum Brain Networks (QBraiNs) as an emerging interdisciplinary endeavour, integrating knowledge and methods from neurotechnology, artificial intelligence (AI), and quantum computing (QC). The objective of QBraiNs is to establish direct communications between the~\textit{human brain} and ~\textit{quantum computers}. We foresee the development of highly connected networks of wetware and hardware devices, processing classical and quantum computing systems, mediated by Brain-Computer Interfaces (BCI) and AI. Such networks will involve unconventional computing systems and new modalities of human-machine interaction.

This paper introduces a first attempt at controlling a qubit with mental activity. We developed a proof-of-concept system, which demonstrates how a person can rotate and measure a qubit using brain signals. 

However, due to limitations imposed by currently available quantum computing hardware and brain sensing technology, our proof-of-concept runs on a software simulation of a quantum computer. Nevertheless, we are one step closer to interfacing the brain with real quantum machines, as improvements in hardware technology at both fronts become available in time to come.

We invented a method to encode neural correlates of mental activity as instructions for a quantum processor. Brain data are read utilising electrodes placed on the scalp of a person, who learns how to produce the required mental activity to issue instructions to rotate and measure a qubit.

By way of previous related work, Kanas \textit{et al.} ~\cite{kanas2014} hinted at the possibility of interfacing the brain with quantum computers. Other speculative propositions were put forward by Pesa and Zizzi ~\cite{Pesa2009} and Musha ~\cite{musha2011}. However forward-thinking as these works may sound, none of them present a concrete experiment or demonstration to support their cases. To the best of our knowledge, the first ever practical demonstration of BCI using quantum computing was reported by Miranda ~\cite{miranda2021}.

The goal of our research is to go a step beyond using quantum computing to analyse brain signals for controlling devices, such as a robot, a vehicle, or a musical instrument, as introduced in ~\cite{miranda2021}. Rather, here we envisage the possibility of forging deeper connections between brains and quantum computers. The ultimate goal is to be able to affect the states of quantum computers with the mind.

\section{Codes of Brain Activity}
\label{sec:codes_brain}

The Homo sapiens' brain is one of the most complex systems known to science. It has circa one hundred billion neurones forming a network of quadrillions of connections ~\cite{Squite2008}. The amount of information that circulates through this network is, although probably bounded, immense. Essentially, neurones are electrical entities. They communicate with one another through action potentials and chemical neurotransmitters. These action potentials are often referred to as {\it spikes}. 

There exists technology nowadays to record neural communication at various levels: from the microscopic level of neurone-to-neurone communication, to higher levels of communication between networks of neurones. Unfortunately, most of this technology is impractical for deployment outside highly specialised research laboratories. At the same time, the engineering to develop sensors made with bioelectronics and nanomaterials is progressing fast to improve this scenario~\cite{Chen2017}. 

Even though sensing technology is becoming increasingly sophisticated, the understanding of the meaning of sensed signals remains very problematic. We may be able to detect neural signals fairly accurately nowadays, but we would not necessarily know what they mean. For instance, it is not hard to render sequences of spikes as sequences of binary numbers for a digital computer to process. But we would have very little clues about what the neurones are communicating to each other. Of course, AI can provide solutions here, as is the case of the machine learning (ML) algorithms for neural decoding introduced in ~\cite{Glasser2020}.

\begin{figure}[htbp]
\begin{center}\vspace{0.2cm}
\includegraphics[width=0.65\linewidth]{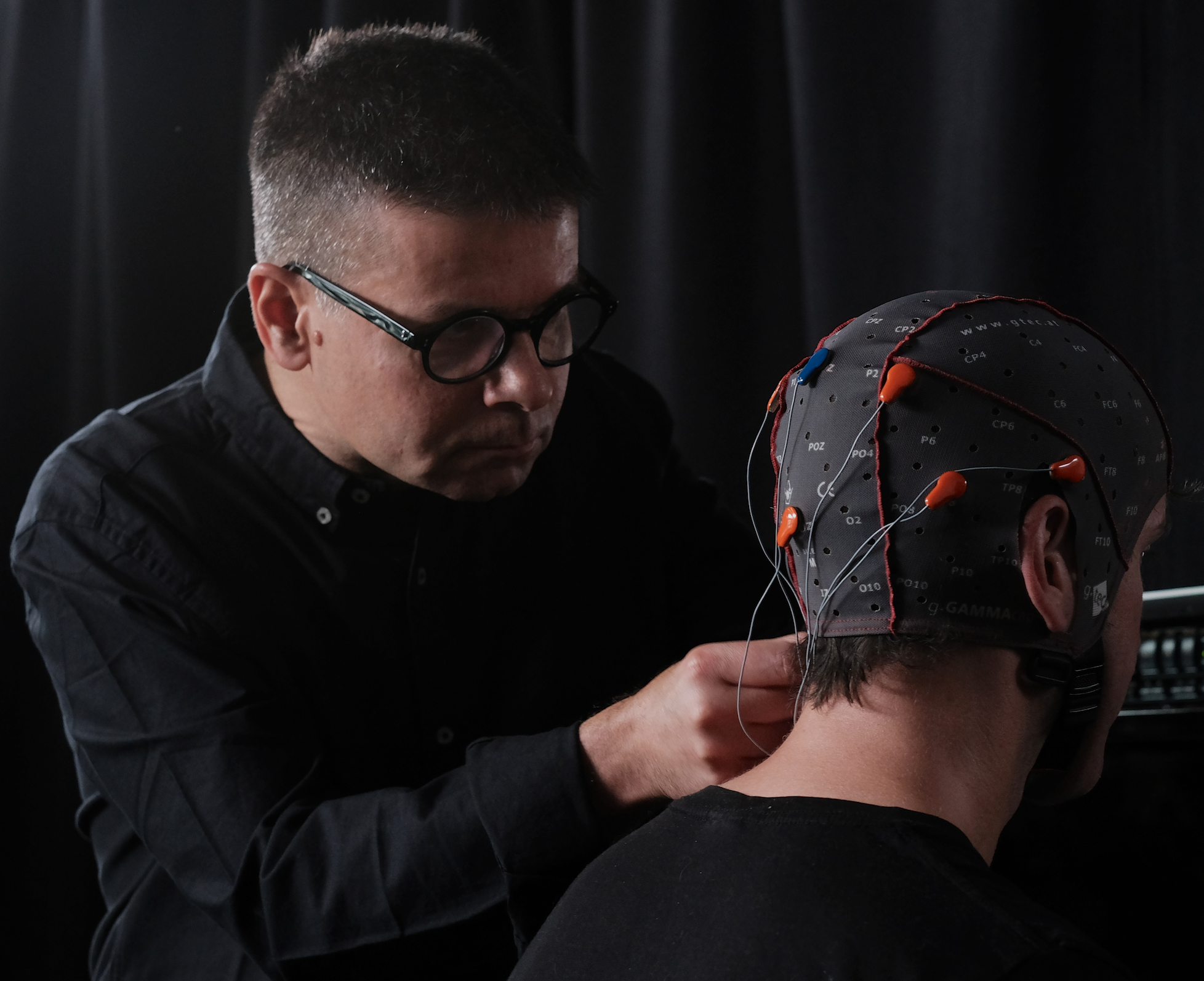}
\caption{Brain activity can be read using electrodes strategically placed on the human scalp.}
\label{fig:EEG_cap}
\end{center}
\end{figure}

A widely used method for reading electrical brain activity is to use electrodes placed on the scalp of a person (Fig.~\ref{fig:EEG_cap}). This recording is called the electroencephalogram, or EEG~\cite{Marcuse2015}. There is a plethora of different EEG recording systems commercially available. They are of varying reliability; the low-cost ones usually relay more spurious signals than actual EEG. It is also possible to record electrical brain activity with electrodes surgically implanted under the skull, on the surface of the cortex, or deep inside the brain; e.g., electrocorticography (ECoG)~\cite{Parvizi2018}. Surgically implanted electrodes provide substantially better signals to work with than scalp electrodes. But brain implants are not routinely used in research with humans at present for obvious health and safety reasons\footnote{Other technologies for brain scanning include functional Magnetic Resonance Imaging (fMRI), near-infrared spectroscopy (NIRS), and magnetoencephalography (MEG). However, these are prohibitively expensive, less portable, and (by the time of writing) offer inadequate time-resolution for BCI purposes.}.

\subsection{The electroencephalogram}

For this project, we adopted scalp EEG. We used an affordable off-the-shelf mid-range device manufactured by g.tec, Graz, Austria\footnote{https://www.unicorn-bi.com/}. It consists of a cap furnished with electrodes and a transmitter that relays the EEG wirelessly to a computer. 

\begin{figure}[htbp]
\begin{center}\vspace{0.2cm}
\includegraphics[width=0.6\linewidth]{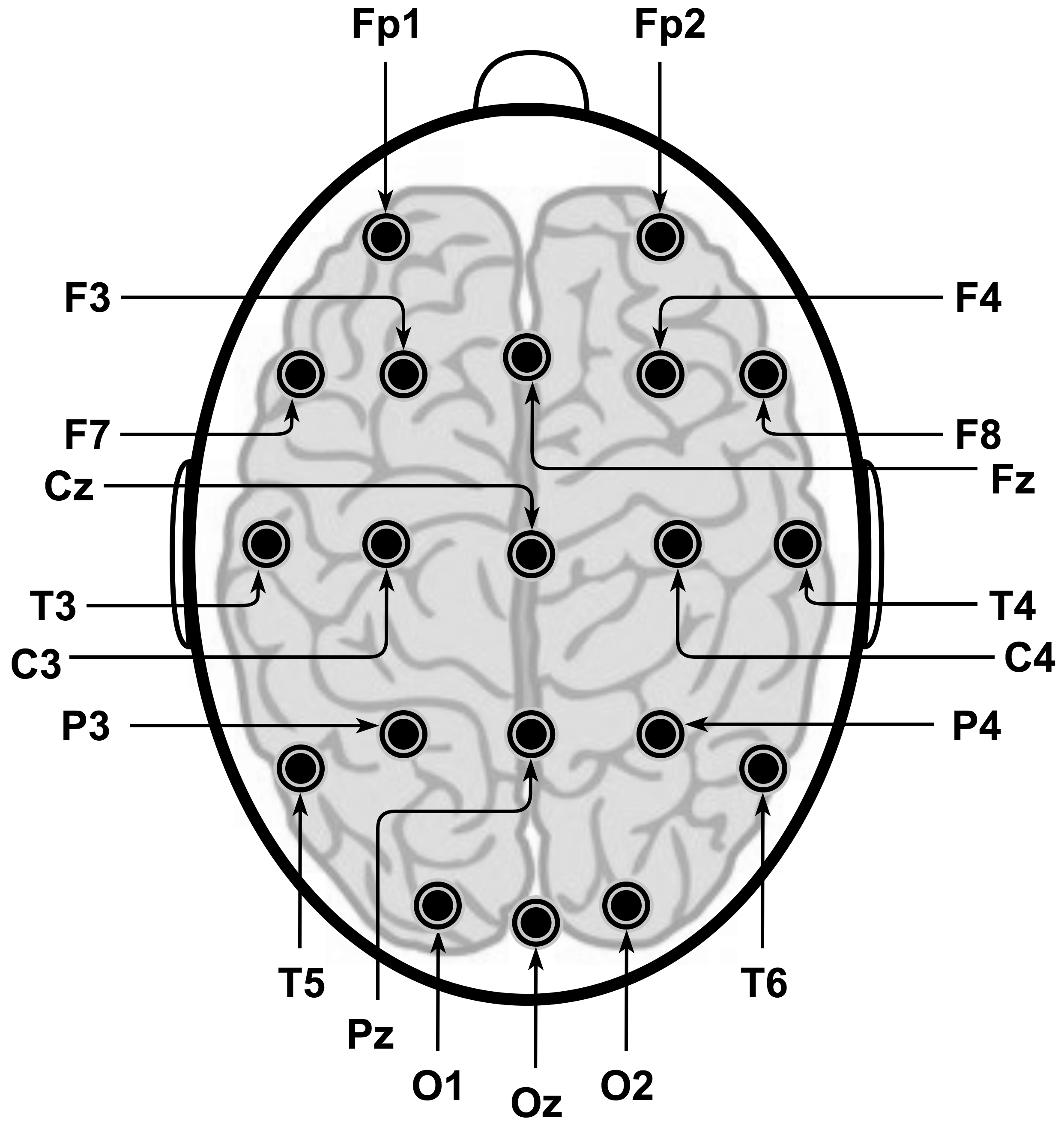}
\caption{Widely used scheme for positioning electrodes for EEG recording.}
\label{fig:EEG_montage}
\end{center}
\end{figure}

The standard scheme for positioning electrodes on the scalp is shown in Fig.~\ref{fig:EEG_montage}. The terminology for referring to the positioning of the electrodes uses letters to indicate a brain region and a number: Fp (pre-frontal), F (frontal), C (central), T (temporal), P (parietal) and O (occipital). Odd numbers are for electrodes on the left side of the head and even numbers for those on the right side; the letter ``z'' stands for the central region. For this project, we used eight electrodes (i.e., eight EEG channels) positioned at F8, F7, Cz, Pz, C4, C3, T6 and T5, respectively.

Power spectrum analysis is a popular method to extract information from EEG. This method breaks the EEG signal into different frequency bands and reveals the distribution of power between them. Power spectrum analysis is widely used in BCI research because it reveals patterns of brain activity that can be recognised automatically and translated into commands for a system. Research exploring the mental correlates of EEG usually considers spectral components up to 40 Hz~\cite{Kropotov2008}. There are four recognised spectral frequency bands, or \textit{EEG rhythms}, each of them associated with specific mental states (Table ~\ref{table:EEG_rhythms}). 

\begin{table}[htbp]
\centering
\begin{tabular}{|c|c|c|}
\hline
\textbf{Frequency Bands} & \textbf{Rhythms} & \textbf{Mental States} \\ \hline
{$f < 4$} & {delta} & {Sleep} \\ \hline
{$4 \leq f <  8$} & {theta} & {Drowsiness} \\ \hline
{$8 \leq f < 15$} & {alpha} & {Low arousal; unfocused; relaxed} \\ \hline
{$15 \leq f < 40$} & {beta} & {High arousal; focused; excited} \\ \hline
\end{tabular}
\caption{\footnotesize{Typical EEG rhythms and associated mental states. Frequency bands are expressed in Hertz (Hz).}}
\label{table:EEG_rhythms}
\end{table}

\subsection{Encoding method}
\label{sec:encoding-method}

We  developed a simple method to encode EEG as instructions to rotate a qubit.  The method takes into account two mental states: low arousal (a.k.a. relaxed) and high arousal (a.k.a. excited). However, to control the qubit, we need at least four different instructions. As the number of instructions is greater than the number of mental states, we sequentially relayed instructions to the system through unique `brain codes'. These are Morse-like binary codes.

As shown in table \ref{table:commands}, there is a unique brain code associated with each instruction, where 0 and 1 correspond to relaxed and excited mental states, respectively. The instructions are as follows:

\begin{itemize}
\item \{0, 1\}:  This is the instruction to start the program, which initializes the connection with the quantum system. None of the other instructions would work without this initialization.
\item \{1, 1\}: This instruction increases the angle of rotation, by a pre-defined amount.
\item  \{0, 0\}: This instruction decreases the angle of rotation, by a pre-defined amount. 
\item  \{1, 0\}: This instruction has two functions. When it occurs for the first time, it changes the axis of rotation on the Bloch sphere (Figure \ref{fig:qubit}), from $z$ (vertical axis) to $y$ (horizontal axis), and vice-versa. Then, when it occurs for the second time, the system measures the qubit.
\end{itemize}

\begin{table}[htbp]
	\centering
\begin{tabular}{|c|c|l|}
	\hline
	\textbf{Command} & \textbf{Brain Code} & \multicolumn{1}{c|}{\textbf{Description}}                                                                                                             \\ \hline
	Start program & \{0, 1\} & \begin{tabular}[c]{@{}l@{}}A relaxed state followed by \\ an excited state. This \\ initializes the connection \\ with the quantum computer.\end{tabular} \\ \hline
	Increase angle & \{1, 1\} & \begin{tabular}[c]{@{}l@{}}An excited state followed by \\ another excited state. This \\ increases the angle of rotation.\end{tabular} \\ \hline
	Decrease angle & \{0, 0\} & \begin{tabular}[c]{@{}l@{}}A relaxed state followed by \\ another relaxed state. This \\ decreases the angle of rotation.\end{tabular}  \\ \hline
	\begin{tabular}[c]{@{}c@{}}Change axis or \\ measure\end{tabular} & \{1, 0\} & \begin{tabular}[c]{@{}l@{}} An excited state followed by \\ a relaxed state. On its first \\ occurrence, it shifts the axis \\of rotation. On its second \\ occurrence, it measures the \\ qubit.\end{tabular}  \\ \hline
\end{tabular}
	\caption{\footnotesize{Different instructions passed to the quantum computer through unique brain codes.}}
	\label{table:commands}
\end{table}

\section{Machine Learning}

Section~\ref{sec:codes_brain} already hinted that the task of establishing what brain signals mean is a fiendish problem. And the fact that the EEG signal is very noisy makes this even more complicated. 

The EEG signal captured by surface electrodes is severely distorted by cortical fluids, the meninges\footnote{These are membranes that envelop the brain.}, the skull, skin and hair; sometimes even the type of shampoo one uses to wash their hair with can cause problems. The signal is unreliable, even to identify only two different classes of mental states. Hence, we use machine learning to harness the capability of the system to identify them.

In order to teach the system to classify between two states of mind, we need to compile a training set with labelled data produced by the user. 

First of all, the system has to be calibrated for the specific user. And this person needs to train themselves how to produce EEG corresponding to relaxed and excited mental states, respectively~\cite{Liebenson2009}. For instance, closing the eyes is one of the easiest and most pragmatic ways to induce the brain to produce (`relaxing') alpha rhythms. Once the user has practised how to achieve a relaxed state of mind, then this is effortlessly achievable with the eyes open. People who are trained in meditation techniques (e.g., yoga) are able to produce alpha rhythms with ease. Beta rhythms can be produced by imagining, or remembering, a stressful situation. Mentally solving a puzzle or a mathematical problem can induce the brain to produce (`exciting') beta rhythms.

Once the user has rehearsed to switch between the two states of mind, then samples of EEG signals corresponding to the respective states are recorded to form a training data set for the classifier. Next, we perform short-time fast Fourier transform (FFT)  analysis on each sample to calculate their average power in the alpha and beta frequency bands. These values are used as features to teach the samples' profiles to a machine learning algorithm.

For the machine learning, we adopted the $k$-Nearest Neighbors (kNN) method. We implemented this using the scikit-learn Python library version 1.0.1\footnote{https://scikit-learn.org/stable/}. kNN is a supervised machine learning method widely used for classification and regression ~\cite{{aha1991}}. In the case of classification, it is  based on assigning a class (or label) to a given sample, to which most of its $k$ neighbours (in a given metric space) belong to. 

The sample data set was split into two subsets, a training and a test set, respectively. The former is used for the calculation of distances (Eq. \ref{eq:eucldistance}). And the latter is used to simulate how the system would work in a real environment.

\begin{equation}
    d_{ij} = ||\mathbf{x_i} - \mathbf{x_j}||
    \label{eq:eucldistance}
\end{equation}

 In particular, kNN calculates the distance from each sample in the test set to the samples of the training set. Then, distances are ordered from smallest to largest, and the $k$ closest samples are selected. The next step involves querying the labels of the selected $k$ samples. As we are dealing with a classification problem, a voting strategy is used to decide the class of the test sample; i.e., the most voted label is used as the selected class.

The degrees of similarity between the samples are calculated using Euclidean distance measurements. The algorithm calculates all possible pairwise Euclidean distances between them. Samples that are close to each other are assigned the same label. Our assumption is that similar brain activities have EEG profiles that are close to each other. Thus, kNN enables the system to determine the label (or, `class') of new incoming EEG data using a distance criterion.

\section{Proof-of-Concept System}
\label{sec:demo}

\begin{figure}[htbp]
	\begin{center}\vspace{0.3cm}
		\includegraphics[width=0.4\linewidth]{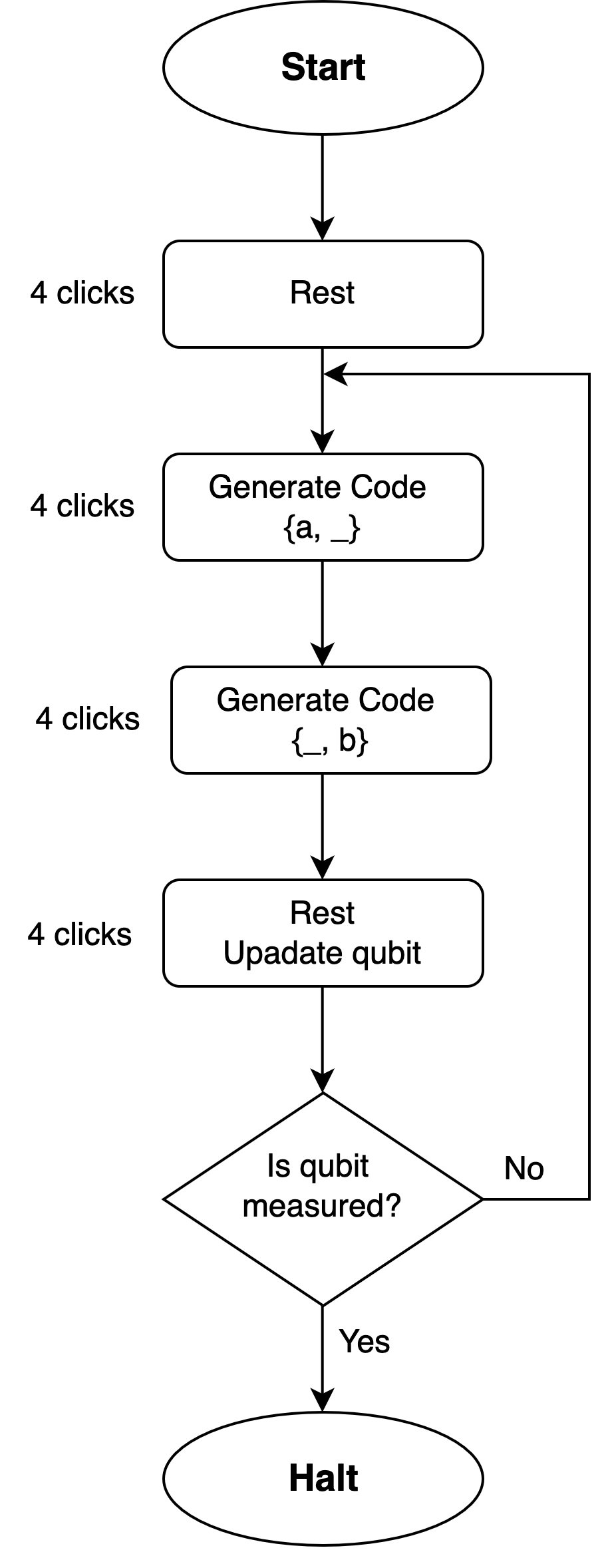}
		\caption{System flow diagram.}
		\label{fig:flow-diagram}
	\end{center}
\end{figure}

As explained in section \ref{sec:encoding-method},  a user alters their mental states to generate brain codes, or instructions, to rotate a qubit. There is a metronome to synchronize the brain with the system. It emits an audible `click' every second. The system builds the brain codes within a window of time lasting for four clicks (i.e., four seconds). The flow diagram in Figure \ref{fig:flow-diagram} illustrates how the system works. 

\begin{figure}[htbp]
	\begin{center}\vspace{0.3cm}
		\includegraphics[width=0.9\linewidth]{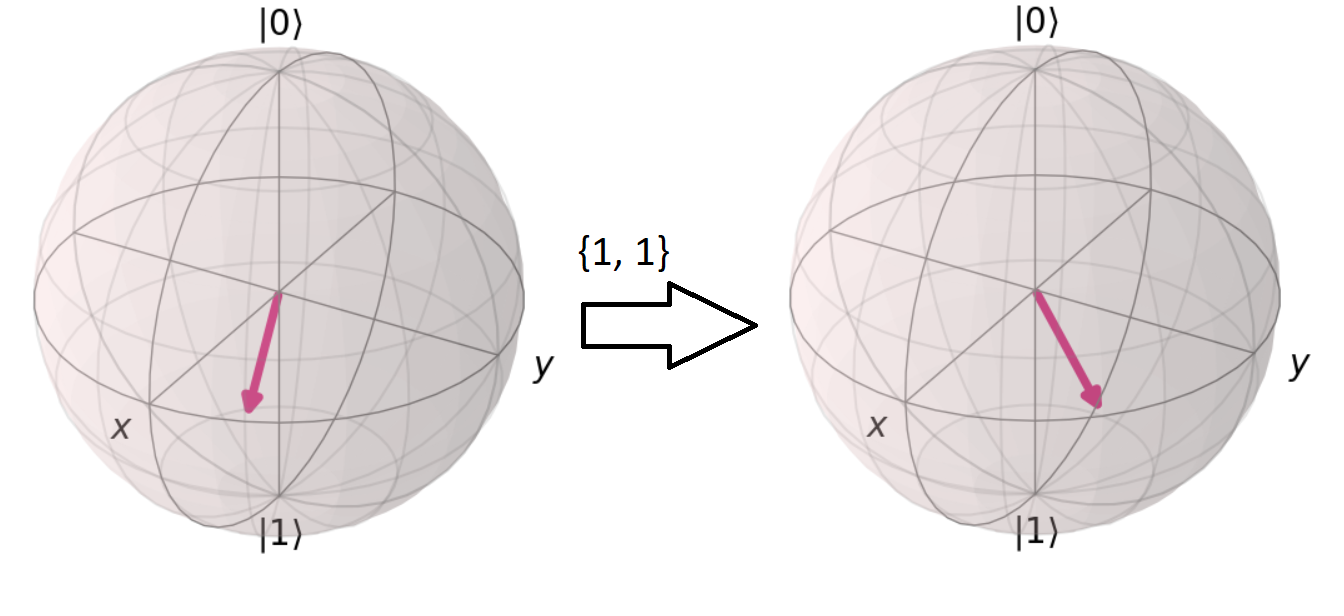}
		\caption{A snapshot of rotating the qubit using the brain code \{1, 1\}.}
		\label{fig:qubit}
	\end{center}
\end{figure}

Initially, the system emits four clicks, which prompts the user to be ready to start a session. Subsequently, the brain activity detected during the following four clicks will correspond to the first digit in the code. Similarly, the second digit is established through the next four clicks. Then, a rest period of four clicks is provided to enable the user to monitor the output; that is, to see if the desired qubit rotation has been achieved. Then, the cycles recommences, and so on. Figure \ref{fig:qubit} shows a snapshot of rotating the qubit with the code \{1, 1\}. In this case, the system detected two consecutive excited mental states in the EEG. This instructed the system to rotate the qubit to the right by a given angle. As a convention, in the context of Figure \ref{fig:qubit}, to `increase the angle of rotation' means to move the state vector to the right side of the image. Conversely, to `decrease the angle of rotation' means to move the state vector to the left.

A video demonstration and programming code are available at the ICCMR GitHub repository: \href{https://github.com/iccmr-plymouth/Quantum-BCI}{https://github.com/iccmr-plymouth/Quantum-BCI}.

\section{Concluding discussion}

\subsection{Towards BCI with quantum hardware}
\label{sec:hardware}

Our proof-of-concept demonstration currently runs on an IBM Quantum simulator\footnote{https://www.ibm.com/quantum-computing/services}. In general, quantum simulators offer more controllability than real quantum computers and, for a small number of qubits, there would be no much difference in performance. Currently, to use a real quantum computer, a program needs to be sent to a machine through a cloud service for batch processing. It is placed in a queue to be computed at a later time. Then, the results are sent back to the client computer. It is not uncommon to having to wait for dozens of minutes until a queued job is processed. This is problematic because our system needs real time access to a qubit.

It is important to note, however, that even if current providers of quantum computing hardware facilities would grant us direct access to their machines, our system would need a specific range of parameters that are not generally available by the time of writing.

However, the fact that we used a simulator of a superconducting quantum processor does not bind our work to superconducting  technology. In fact, our system would not work well on superconducting quantum devices as we know them today. The caveat is that the operational timescales of EEG and superconducting qubits are orders of magnitude apart, ranging from seconds (in the EEG domain), to microseconds (in the qubits domain). In simpler words, the machine would need to maintain qubits coherent for a prohibitively long time until the brain produces a command. To rotate and measure a qubit directly with human brain activity, we would need a huge leap in qubit coherence time that cannot be afforded with current superconducting technology\footnote{Coherence time is the length of time a qubit is able to hold quantum information. This requires physical qubits to remain highly isolated from the surrounding environment. When a qubit is disrupted by external interference (e.g., background noise from vibrations, temperature changes or stray electromagnetic fields) information about the state of that qubit is destroyed, in a process known as decoherence. This can ruin the ability to exploit quantum mechanics for computation. Longer coherence times enable more quantum operations to be utilised before this occurs.}. It might be possible to alleviate this problem with an operational system that would facilitate countless classical-quantum iterations. 

Fortunately, there are signs that quantum hardware platforms that would be suitable for the types of systems that we are interested in developing are already emerging in a number of academic research labs. For instance, qubits built with spin pairs in diamond~\cite{Bartling} seem to hold coherence for over one minute, which would account for sequences of our brain-generated rotation commands.  At least theoretically, this specification matches the conditions required for our proof-of-concept (Eq. \ref{eq:equa}), even before the optimisations discussed in section \ref{sec:beyondEEG} below.

\begin{eqnarray}
t_{\rm rot} < t_{\rm step} << t_{\rm coh}
\label{eq:equa}
\end{eqnarray}

\noindent
where $t_{\rm rot}$ is the time it takes to produce a finite rotation in the qubit, $t_{\rm step}$ is the time of each step realized by our protocol, while $t_{\rm coh}$ is the total coherent time of the qubit. Moreover, those qubits operate at room temperature~\cite{PrivateComm}, which, on the long run, may enable the manufacturing of more accessible workstations.

\subsection{Looking Beyond EEG}
\label{sec:beyondEEG}

Currently, the system takes four seconds to analyse and classify the EEG signal to generate a digit for our brain code. Thus, it needs eight seconds to compute a code. Obviously, this is a far cry from ideal. There definitely are signal processing techniques ~\cite{paszkiel2020} \cite{im2018} and other robust classification methods to optimise this ~\cite{kolosova2021} \cite{hastie2017}. 

Moreover, EEG correlates of states of mind other than the EEG rhythms listed in Table \ref{table:EEG_rhythms} have been harnessed for BCI; e.g., evoked potential ~\cite{Miranda2011} and motor imagery ~\cite{Padfield2019}. Thus, there are additional alternatives to be explored.

For this project we used an affordable off-the-shelf mid-range EEG device, using surface dry electrodes. There is EEG technology that offers much higher fidelity than the fidelity offered by our equipment. And electrodes surgically implanted under the skull capture considerably better EEG signals than surface ones. 

Furthermore, brain scanning technology that offers more precision than EEG, but which until recently were deemed unsuitable for BCI, are important avenues to be explored. These include Magnetic Resonance Imaging (MRI) ~\cite{huettel2014} and Magnetoencephalography MEG (MEG) ~\cite{hansen2010}. For instance, emerging wearable scanners based on non-cryogenic OPM-MEG\footnote{Optically Pumped Magnetometers MEG.} are promising new devices ~\cite{hilletal2019}. OPM-MEG technology uses {\it{quantum sensors}} to measure magnetic fields generated by electrical activity within the brain.

Nevertheless, in addition to improving brain scanning fidelity and resolution, and the timing scale discrepancy mentioned above (section \ref{sec:hardware}), we need to further develop meaningful brain encoding schemes to communicate with quantum states. 

As a starting point, we proposed a Morse-like binary coding informed by the  way in which the brain functions at all levels. Excitatory and inhibitory processes pervade the functioning of our brain, from the microscopic level of neurones communicating with one another, to the macroscopic level of interaction between larger networks of millions of neurones. The encoding method introduced in \ref{sec:encoding-method} works at the abstract level of EEG rhythms: think of a high arousal (`exciting') mental state as an excitatory neural process and low arousal (`relaxing') as an inhibitory one. 

A better understanding of the meaning of the spiking behaviour of neurones, and networks thereof, plus the ways in which we might be able to control them - voluntarily or involuntarily - are {\it{sine qua non}} for progressing with our approach to interfacing the brain with quantum computers.

\section{Acknowledgements}
The authors acknowledge support from the Spanish government (grant references MCIU/AEI/FEDER PID2019-104002GB-C21 and PID2019-104002GB-C22), the Regional Government of Andalusia, Spain (grant reference P20-00617) and Shanghai's Municipality, China (grant references 2019SHZDZX01-ZX04 and 20DZ2290900).  We are indebted to the University of Plymouth, UK, for access to its brain imaging resources, and the QuTune Project, sponsored by UK's QCS Hub.

\end{document}